\documentclass[sigconf]{acmart}
\usepackage{hyperref}
\usepackage[linesnumbered,ruled,vlined]{algorithm2e}
\usepackage{amsmath}
\usepackage{color}
\usepackage{tabularray}
\usepackage{subcaption}
\usepackage{subcaption}
\usepackage{float} 
\usepackage{hyperref}
\AtBeginDocument{%
  }

\copyrightyear{2025}
\acmYear{2025}
\setcopyright{cc}
\setcctype{by}
\acmConference[RecSys '25]{Proceedings of the Nineteenth ACM Conference on Recommender Systems}{September 22--26, 2025}{Prague, Czech Republic}
\acmBooktitle{Proceedings of the Nineteenth ACM Conference on Recommender Systems (RecSys '25), September 22--26, 2025, Prague, Czech Republic}\acmDOI{10.1145/3705328.3748029}
\acmISBN{979-8-4007-1364-4/2025/09}
\begin{document}

%%
%% The "title" command has an optional parameter,
%% allowing the author to define a "short title" to be used in page headers.
% \title{Estimating Quantum Execution Requirements for Feature Selection in Recommender Systems Using Extreme Value Theory}
\title[Q-EVT: Quantum Execution Estimation for Feature Selection]{Estimating Quantum Execution Requirements for Feature Selection in Recommender Systems Using Extreme Value Theory}
%%
%% The "author" command and its associated commands are used to define
%% the authors and their affiliations.
%% Of note is the shared affiliation of the first two authors, and the
%% "authornote" and "authornotemark" commands
%% used to denote shared contribution to the research.
\author{Jiayang Niu}
\affiliation{%
  \institution{School of Computing Technologies, RMIT University}
  \city{Melbourne}
  \country{Australia}}
\email{s4068570@student.rmit.edu.au}

\author{Qihan Zou}
\affiliation{%
  \institution{School of Mathematics and Statistics, University of Melbourne}
  \city{Melbourne}
  \country{Australia}}
\email{qihanzou@gmail.com}

\author{Jie Li}
\affiliation{%
  \institution{School of Computing Technologies, RMIT University}
  \city{Melbourne}
  \country{Australia}}
\email{hey.jieli@gmail.com}

\author{Ke Deng}
\email{ke.deng@rmit.edu.au}
\affiliation{%
  \institution{School of Computing Technologies, RMIT University}
  \city{Melbourne}
  \country{Australia}}
\email{ke.deng@rmit.edu.au}

\author{Mark Sanderson}
\affiliation{%
  \institution{School of Computing Technologies, RMIT University}
  \city{Melbourne}
  \country{Australia}}
\email{mark.sanderson@rmit.edu.au}

\author{Yongli Ren}

\affiliation{%
  \institution{School of Computing Technologies, RMIT University}
  \city{Melbourne}
  \country{Australia}}
\email{yongli.ren@rmit.edu.au}

%%
%% By default, the full list of authors will be used in the page
%% headers. Often, this list is too long, and will overlap
%% other information printed in the page headers. This command allows
%% the author to define a more concise list
%% of authors' names for this purpose.
\renewcommand{\shortauthors}{Jiayang Niu et al.}

%%
%% The abstract is a short summary of the work to be presented in the
%% article.
\begin{abstract}
Recent progress in quantum computing has advanced research in quantum-assisted information retrieval and recommender systems, especially for feature selection via Quadratic Unconstrained Binary Optimization (QUBO). However, while existing work primarily focuses on effectiveness and efficiency, However, it often neglects the inherent noise and probabilistic nature of quantum hardware. In this paper, we propose a method based on Extreme Value Theory (EVT) to estimate the number of quantum executions (shots) needed to reliably obtain high-quality solutions—comparable to or better than classical baselines. Experiments on both simulators and two physical quantum devices demonstrate that our method effectively estimates the number of required runs to obtain satisfactory solutions on two widely used benchmark datasets.
\end{abstract}

%%
%% The code below is generated by the tool at http://dl.acm.org/ccs.cfm.
%% Please copy and paste the code instead of the example below.
%%
\begin{CCSXML}
<ccs2012>
    <concept>
        <concept_id>10002951.10003317.10003347.10003350</concept_id>
        <concept_desc>Information systems~Recommender systems</concept_desc>
    <concept_significance>300</concept_significance>
    </concept>
   <concept>
       <concept_id>10010583.10010662.10010668</concept_id>
       <concept_desc>Computer systems organization~Quantum computing</concept_desc>
       <concept_significance>500</concept_significance>
   </concept>
   <concept>
       <concept_id>10002950.10003648.10003662</concept_id>
       <concept_desc>Mathematics of computing~Probability and statistics</concept_desc>
       <concept_significance>300</concept_significance>
   </concept>
</ccs2012>
\end{CCSXML}
\ccsdesc[500]{Mathematics of computing~Probability and statistics}
\ccsdesc[500]{Information systems~Recommender systems}
\ccsdesc[500]{Computer systems organization~Quantum computing}

%%
%% Keywords. The author(s) should pick words that accurately describe
%% the work being presented. Separate the keywords with commas.
\keywords{Quantum computing, Recommender systems, Feature selection, Extreme value theory}

%\received{20 February 2007}
%\received[revised]{12 March 2009}
%\received[accepted]{5 June 2009}

%%
%% This command processes the author and affiliation and title
%% information and builds the first part of the formatted document.
\maketitle

\section{Introduction}
Recent advancements in quantum technology, along with growing media attention toward commercial quantum computers, suggest that qubit-based hardware may not only complement but in certain domains may outperform traditional transistor-based computing systems that have dominated for decades~\cite{king2025beyond, ai2024quantum}. Researchers have begun to explore how these quantum techniques can benefit Information Retrieval and Recommender Systems (RecSys). For instance, Nembrini et al. \cite{ferrari2022towards} embedded several classical feature selection strategies into Quadratic Unconstrained Binary Optimization (QUBO) formulations, while Niu et al.~\cite{niu2024performance} applied counterfactual analysis to assess individual feature contributions to model performance and integrated these insights into their QUBO model. However, these studies primarily focus on the effectiveness and the efficiency when using quantum annealers (e.g., D-Wave), leaving two key \emph{research gaps}: \textbf{(1)} Because quantum measurements are inherently probabilistic, it is necessary to run many shots\footnote{A shot refers to a single execution of a quantum program, including measurement. Multiple shots are used to sample from the output distribution.}, so as to determine the most likely or optimal results. For a fixed problem (e.g., feature selection in RecSys), it remains unclear how many shots are required to obtain high-quality results. \textbf{(2)} Compared to quantum annealers, gate-based quantum computers are more powerful and versatile, as they support a broader class of quantum algorithms beyond optimisation. Thus, exploring their use in this domain holds significant potential. 

In this paper, we aim to answer the research question: 
\emph{How to estimate the number of required shots for gate-based quantum computers to achieve good results in feature selection for recommendation tasks?} 
Specifically, we propose Quantum-EVT, a practical and theoretically grounded framework based on Extreme Value Theory (EVT)~\cite{kotz2000extreme} for estimating the required number of quantum executions (shots), which provides confidence-based estimates that predict when quantum solvers are likely to outperform classical baseline model in terms of energy optimisation. Following existing work~\cite{nembrini2021feature, ferrari2022towards, niu2024performance}, we frame the feature selection problem in recommendation as a QUBO formulation for a click-through-rate prediction task and solve it using the Quantum Approximate Optimisation Algorithm (QAOA)~\cite{farhi2014quantum}, which is designed to solve combinatorial problems on a gate-based quantum device. However, there are two main challenges: \textbf{(1)} Deploying EVT requires several key statistical parameters, which must be estimated from many quantum executions to obtain some extreme value samples. While more samples generally lead to better results, there is a trade-off with computational cost. Thus, determining the minimum sufficient number of extreme samples becomes essential. \textbf{(2)} The quantum energy values derived from QUBO problems are inherently discrete, while EVT assumes a continuous distribution. Therefore, we must adapt the discrete quantum outputs to be compatible with EVT. To overcome these challenges, \textbf{(1)} we propose an algorithm that statistically determines the minimal number of extreme samples needed for stable EVT parameter estimation; \textbf{(2)} we introduce a technique to approximate a continuous distribution by adding uniform noise to the discrete quantum energy values. 
% We summarize our main contributions as follows: 
% \begin{itemize}
% \item An EVT-based framework for estimating the required number of quantum executions, providing confidence-based shot estimates that predict when quantum solvers outperform classical simulated annealing (SA).

% \item The first application of gate-based quantum computers to solve the feature selection problem in recommender systems. 

% \item A comprehensive evaluation of EVT-based predictions, demonstrating that the estimated number of runs is both accurate and robust under realistic settings, which also validates EVT as a robust tool for modelling quantum optimisation behaviour.
% \end{itemize}

The main contributions of this work are: 
\textbf{(i)} An EVT-based framework for estimating the required number of quantum executions, providing confidence-based shot estimates that predict when quantum solvers outperform classical simulated annealing (SA).
\textbf{(ii)} The first application of gate-based quantum computers to solve the feature selection problem in recommender systems. 
\textbf{(iii)} A comprehensive evaluation of EVT-based predictions, demonstrating that the estimated number of runs is both accurate and robust under realistic settings, which also validates EVT as a robust tool for modelling quantum optimisation behaviour.

\section{RELATED WORKS}
Recent studies have explored the use of quantum optimization, particularly Quadratic Unconstrained Binary Optimization (QUBO), for feature and instance selection in information retrieval~\cite{pasin2024quantum} (IR) and recommender systems (RecSys). CQFS~\cite{nembrini2021feature}, MIQUBO, CoQUBO, QUBO-Boosting~\cite{ferrari2022towards}, and PDQUBO~\cite{niu2024performance} all formulate the feature selection task as a QUBO problem by designing different strategies for constructing the coefficient matrix~$Q$. These strategies range from leveraging collaborative signals~\cite{nembrini2021feature}, to statistical measures such as mutual information and correlation~\cite{ferrari2022towards}, to performance-oriented counterfactual analysis~\cite{niu2024performance}. These works also emphasize that quantum annealers, particularly D-Wave systems, are well-suited to solving QUBO problems due to their hardware design. However, PDQUBO additionally notes that annealer performance deteriorates as problem size increases, revealing stability concerns. Zaborniak et al.~\cite{zaborniak2021benchmarking} also quantitatively benchmark the Hamiltonian noise on different generations of D-Wave hardware, showing significantly higher noise levels in newer systems (e.g., Advantage\_system1.1) compared to earlier models.

Unlike quantum annealers, gate-based quantum computers are universal devices capable of running a broader class of quantum algorithms. The Quantum Approximate Optimization Algorithm (QAOA)~\cite{farhi2014quantum} is a prominent candidate for solving QUBO problems on such devices. QAOA leverages parameterized variational circuits to approximate optimal solutions, enabling hybrid quantum-classical optimization pipelines.  To complement the empirical use of QAOA, we adopt \textit{Extreme Value Theory} (EVT)~\cite{coles2001introduction, gomes2015extreme} as a principled statistical framework to model the distribution of low-energy outcomes obtained from repeated quantum executions. EVT naturally fits the structure of quantum optimization, allowing us to estimate how many repetitions are required to reach a target solution quality (e.g., matching classical baselines) under a specified confidence level.

\section{PRELIMINARIES}

\textbf{Quadratic Unconstrained Binary Optimization (QUBO)} is a mathematical framework for representing binary optimization problems. It optimizes a binary vector $\mathbf{x} \in \{0,1\}^n$ (with the $i$-th entry $x_i$ denoting the $i$-th feature) over a symmetric matrix $Q \in \mathbb{R}^{n \times n}$:
\begin{equation}
\label{eq:QUBO_penalty}
\min_{\mathbf{x}} Y = \mathbf{x}^{\top} Q \mathbf{x} + \left(\sum_{i=1}^{n} x_i - k\right)^2, \quad \mathbf{x} \in \{0,1\}^n
\end{equation}
where $Y$ is the ground state energy in quantum solutions, and the penalty term $\left(\sum_{i} x_i - k\right)^2$ enforces a cardinality constraint, ensuring that exactly $k$ features are selected. This constraint-augmented QUBO formulation is particularly well-suited for tasks such as feature selection~\cite{ferrari2022towards}, where selecting a fixed-size subset of features is required. 
The resulting energy $Y$ from QUBO optimization serves as a proxy for the quality of the selected subset—the lower the energy, the better the expected recommendation performance.

% \textbf{Quantum Approximate Optimization Algorithm (QAOA)}~\cite{farhi2014quantum} is a hybrid quantum-classical method designed for solving combinatorial optimization problems on gate-based quantum devices. To apply QAOA to a QUBO problem, binary feature $x_i \in \{0,1\}$ are first mapped to spin variables $z_i \in \{-1,1\}$ via $x_i = \frac{1 + z_i}{2}$. This transformation converts QUBO objective into an Ising Hamiltonian composed of single-qubit ($Z_i$) and two-qubit ($Z_i Z_j$) terms.
% QAOA constructs a parameterized quantum state by alternately applying two unitaries: a cost operator $U_C(\gamma) = e^{-i\gamma H_C}$, where $H_C = \sum_i h_i Z_i + \sum_{i<j} J_{ij} Z_i Z_j$ encodes the problem structure, and a mixer operator $U_M(\beta) = e^{-i\beta H_M}$, where $H_M = \sum_i X_i$ enables state exploration. These operations are applied in $p$ layers to an initial uniform superposition $|\psi_0\rangle = H^{\otimes n} X^{\otimes n}|0\rangle^{\otimes n}$.
% The variational parameters $(\gamma, \beta)$ are optimized classically to minimize the expected energy $\langle \psi(\gamma, \beta) | H_C | \psi(\gamma, \beta) \rangle$. Low-energy measurement outcomes correspond to better approximate solutions to the QUBO problem.

\textbf{Quantum Approximate Optimization Algorithm (QAOA)}~\cite{farhi2014quantum} is a hybrid quantum-classical method designed for solving combinatorial optimization problems on gate-based quantum devices. To apply QAOA to a QUBO problem, binary variables $x_i \in \{0,1\}$ are first mapped to spin variables $z_i \in \{-1,1\}$ via $x_i = \frac{1 + z_i}{2}$. Substituting this into the QUBO formulation yields an equivalent objective expressed entirely in terms of $z_i$:

\begin{equation}
\label{eq:QUBO_to_Ising}
\begin{aligned}
\min_\mathbf{x} Y &= \frac{1}{4}\sum_{i,j} Q_{ij}(1 + z_i)(1 + z_j) 
+ \frac{1}{4}\left(\sum_{i=1}^{n} (1 + z_i) - 2k\right)^2, \\
&= \frac{1}{2}\sum_{i=1}^{n}\left(\sum_{j=1}^{n} Q_{ij} + n - 2k\right)z_i 
+ \frac{1}{4}\sum_{i,j=1}^{n}\left(Q_{ij} + 1\right)z_i z_j 
\end{aligned}
\end{equation}

This transformation converts the QUBO objective into an Ising Hamiltonian composed of single-qubit ($Z_i$) and two-qubit ($Z_i Z_j$) terms. QAOA constructs a parameterized quantum state by alternately applying two unitaries: a cost operator $U_C(\gamma) = e^{-i\gamma H_C}$, where $H_C = \sum_i h_i Z_i + \sum_{i<j} J_{ij} Z_i Z_j$ encodes the problem structure, and a mixer operator $U_M(\beta) = e^{-i\beta H_M}$, where $H_M = \sum_i X_i$ enables state exploration. These operations are applied in $p$ layers to an initial uniform superposition $|\psi_0\rangle = H^{\otimes n} X^{\otimes n}|0\rangle^{\otimes n}$. The variational parameters $(\gamma, \beta)$ are optimized classically to minimize the expected energy $\langle \psi(\gamma, \beta) | H_C | \psi(\gamma, \beta) \rangle$. Low-energy measurement outcomes correspond to better approximate solutions to the QUBO problem.

\textbf{PDQUBO}~\cite{niu2024performance} formulates feature selection as a QUBO problem by encoding the impact of individual and joint feature removals on model performance into the matrix $Q$ in Eq.~\ref{eq:QUBO_penalty}. Specifically, the diagonal and off-diagonal elements of $Q$ are computed as: $Q_{ij} = G(\mathcal{F}) - G(\mathcal{F}_{\text{mask}}^{i})$ when $i=j$; otherwise $Q_{ij} = G(\mathcal{F}) - G(\mathcal{F}_{\text{mask}}^{ij})$, 

where $G(\cdot)$ is the performance metric (e.g., Area Under the Curve - AUC) of a trained recommendation model. $\mathcal{F}_{\text{mask}}^i$ and $\mathcal{F}_{\text{mask}}^{ij}$ denote feature subsets with one or two features removed. Thus, $Q$ reflects both individual feature relevance and pairwise interactions. 

% \textbf{PDQUBO}~\cite{niu2024performance} formulates feature selection as a QUBO problem by encoding the impact of individual and joint feature removals on model performance into the matrix $Q$ in Eq.~\ref{eq:QUBO_penalty}, which is computed as:

% \begin{equation}
% \label{eq:pdqubo}
% Q_{ij} = \begin{cases}
% G(\mathcal{F}|\Theta)_{\text{Mtc}} - G(\mathcal{F}_{\text{mask}}^{i}|\Theta)_{\text{Mtc}}, & \text{if } i=j,\\[6pt]
% G(\mathcal{F}|\Theta)_{\text{Mtc}} - G(\mathcal{F}_{\text{mask}}^{ij}|\Theta)_{\text{Mtc}}, & \text{if } i\neq j
% \end{cases}
% \end{equation}

% where $G(\mathcal{F}|\Theta)_{\text{Mtc}}$ denotes the performance metric (e.g., AUC) obtained by training the recommendation model (e.g., DeepFM or FiBiNET) using the full feature set $\mathcal{F}$. The sets $\mathcal{F}_{\text{mask}}^{i}$ and $\mathcal{F}_{\text{mask}}^{ij}$ represent feature subsets obtained by masking (removing) one or two features, respectively. Therefore, diagonal elements $Q_{ii}$ reflect the individual importance of each feature, while off-diagonal elements $Q_{ij}$ capture joint interactions between pairs of features. 

% The final objective value (energy $Y$) in PDQUBO, as obtained by solving the QUBO problem, directly indicates the quality of the selected feature subset. A lower energy value signifies a better-performing feature subset in the recommendation task.

\section{Quantum-EVT (Q-EVT)}

\begin{figure}[h]
    \centering
\includegraphics[width=1.0\linewidth]{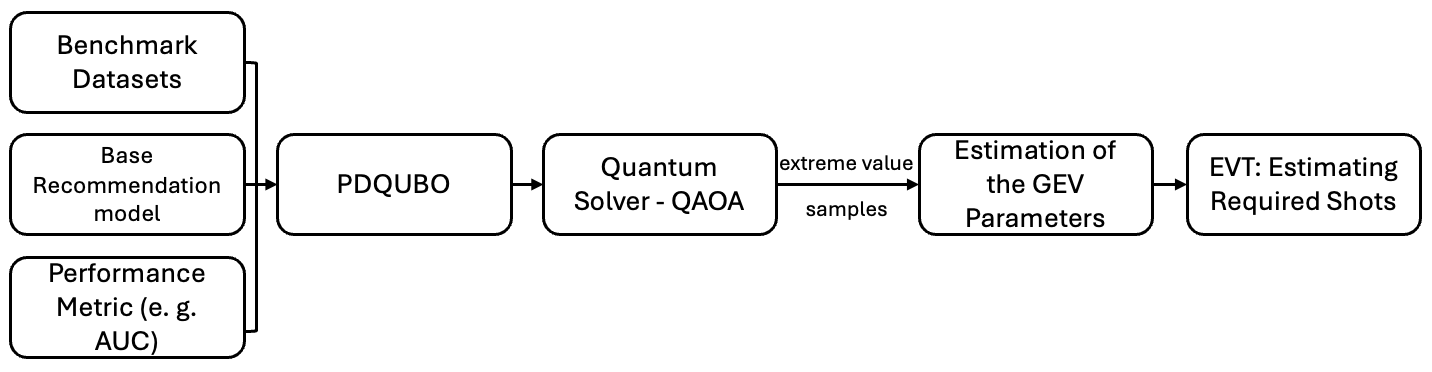}
    \caption{The Overview of Q-EVT}
    \label{fig:qevt}
\end{figure}

We propose Q-EVT (as shown in Figure~\ref{fig:qevt}), a framework for estimating the number of quantum executions (shots) required for a given task. It formulates feature selection in recommendation as a QUBO problem, solved using QAOA to obtain extreme value samples for applying Extreme Value Theory (EVT). Q-EVT outputs confidence-based estimates predicting when quantum solvers are likely to outperform classical models in energy optimisation.

\subsection{Extreme Value Theory (EVT)}

QAOA yields probabilistic outputs, where each execution (\textit{shot}) returns only one candidate solution. Due to noise and inherent randomness, optimal solutions are obtained only with limited probability, motivating the use of repeated shots. Extreme Value Theory (EVT)~\cite{kotz2000extreme, kratz2019introduction,smith1990extreme,smith1994extreme} offers a natural framework to analyze the minimum energy values obtained across these repeated executions. By modeling the distribution of such extreme outcomes, EVT enables principled estimation of quantum resources required to achieve desired solution quality.

Assume energy samples $\{y_1, y_2, \dots, y_n\}$ from repeated quantum executions are independent and identically distributed (i.i.d.) from an unknown distribution $F$. EVT primarily analyzes the distribution of block maxima~\cite{haan2006extreme,gomes2015extreme}:
\begin{equation}
M_n = \max(y_1, y_2, \dots, y_n).
\end{equation}

Under certain conditions, the normalized block maximum $M_n$ converges to a Generalized Extreme Value (GEV) distribution~\cite{haan2006extreme}:
\begin{equation}
\mathbb{P}\left( \frac{M_n - \mu}{\sigma} \leq z \right) \to G(z) =
\begin{cases}
\exp\left( -\left[1 + \xi \frac{z - \mu}{\sigma}\right]^{-1/\xi} \right), & \xi \ne 0, \\
\exp\left( -\exp\left(-\frac{z - \mu}{\sigma}\right) \right), & \xi = 0,
\end{cases}
\end{equation}
where $\mu \in \mathbb{R}$ is the location parameter, $\sigma > 0$ is the scale parameter, and $\xi \in \mathbb{R}$ is the shape parameter. While the original formulation focuses on maxima, analyzing minima can be achieved by negating observations ($-y_i$).
However, accurately applying EVT demands reliable estimation of GEV distribution parameters.

\subsection{Estimating Required Extreme Samples}
\label{sec:eres}

The estimation of the GEV parameters ($\mu$, $\sigma$, and $\xi$) requires a sufficient number of extreme value samples. While larger sample sizes improve the stability of parameter estimates, practical constraints—such as the cost and latency of quantum executions necessitate a careful trade-off. Thus, identifying the minimal number of samples required for reliable estimation becomes essential. To address this, we follow the previous work by~\cite{cai2010minimum} (that provided an estimation method for minimum samples with return level as the objective with univariate Shapiro-Wilk normality test) and the idea of the bootstrap method~\cite{efron1994introduction, davison1997bootstrap} to design a statistical procedure that estimates the smallest sufficient sample size $n_{\text{estimate}}$. Starting with a large set of extreme values $Y_{\text{sim}}$ (Each value in $Y_{\text{sim}}$ is a block-wise extreme like $M_n$, collected from repeated quantum executions) obtained from quantum simulations, we compute a stable reference GEV parameter set $\theta_{\text{sim}} = (\mu_{\text{sim}}, \sigma_{\text{sim}}, \xi_{\text{sim}})$. Then, for a range of candidate sample sizes $n \in [n_{\min}, N_{\max}]$, we repeatedly draw random subsets from $Y_{\text{sim}}$, estimate GEV parameters for each subset, and perform two statistical tests on the resulting estimates: Hotelling's $T^2$ test~\cite{anderson1958introduction} to check for consistency with the reference mean and the multivariate Shapiro-Wilk test~\cite{villasenor2009generalization} to assess normality.

The average $p$-values from these tests indicate the stability of parameter estimation at each $n$. The smallest $n$ for which both tests yield $p$-values exceeding 0.05 is selected as $n_{\text{estimate}}$. \textit{Note.:} $N_{\max}$ should be chosen conservatively. If the subset size approaches the total pool $N$, the samples become nearly identical due to the limited size of discrete quantum outcomes. This causes the multivariate Shapiro-Wilk test to fail, as the sampled distributions collapse to a point. While assuming continuous distributions could mitigate this, such an assumption is generally unjustified given the discrete nature of quantum outputs. The complete procedure is summarized in Algorithm~\ref{alg:extreme}.

\begin{algorithm}[h]
\caption{Estimate Required Extreme Samples}
\label{alg:extreme}
\SetKwInOut{Input}{Input}
\SetKwInOut{Output}{Output}
\Input{
  $Y_{\text{sim}}$ --- sequence of simulated extreme values\\
  $n_{\min}, N_{\max}$ --- range of sample sizes to test\\
  $I, J$ --- number of repetitions per estimation\\
  $\theta_{\text{sim}} = (\mu_{\text{sim}}, \sigma_{\text{sim}}, \xi_{\text{sim}})$ --- reference GEV parameters
}
\Output{
  $n_{\text{estimate}}$ --- estimated number of extreme samples required
}
\BlankLine
\For{$n = n_{\min}$ \KwTo $N_{\max}$}{
  \For{$j = 1$ \KwTo $J$}{
    \For{$i = 1$ \KwTo $I$}{
      Draw $n$ samples $Y_{ijn}$ from $Y_{\text{sim}}$ with replacement\;
      
      Estimate GEV parameters from $Y_{ijn}$: $\hat{\theta}_{ijn} = (\hat{\mu}_{ijn}, \hat{\sigma}_{ijn}, \hat{\xi}_{ijn})$\;
    }
    Let $\hat{\theta}_{jn}$ denote the collection of estimates for $i=1,\dots,I$\;
    
    Perform Hotelling's $T^2$ test for the null hypothesis 
    \[
    H_0: E(\hat{\theta}_{jn}) = \theta_{\text{sim}},
    \]
    and obtain the p-value $p_{jn,ht2}$\;
    
    Perform the multivariate Shapiro-Wilk normality test on $\hat{\theta}_{jn}$ and obtain the p-value $p_{jn,mst}$\;
  }
  Compute the average p-values: 
  \[
  \bar{p}_{n,ht2} = E(p_{jn,ht2}) \quad \text{and} \quad \bar{p}_{n,mst} = E(p_{jn,mst})
  \]
  over $j = 1,\dots,J$\;
}
Fit regression lines to the pairs $(n, \bar{p}_{n,ht2})$ and $(n, \bar{p}_{n,mst})$\;
Determine the smallest $n$ such that both $\bar{p}_{n,ht2}$ and $\bar{p}_{n,mst}$ exceed $0.05$ based on the intersection of regression lines and $p = 0.05$, and let these be $n_{ht2}$ and $n_{mst}$ respectively\;
\If{$n_{ht2} < n_{mst}$}{
    $n_{\text{estimate}} \gets n_{mst}$ \tcp*[h]{Hotelling's $T^2$ test requires normality}
}
\Else{
    $n_{\text{estimate}} \gets n_{ht2}$
}
\Return{$n_{\text{estimate}}$}\;
\end{algorithm}

\subsection{Estimation of the Required Shots}

Quantum energy samples from QAOA are inherently discrete due to the binary nature of solution vectors, whereas EVT assumes continuous distributions. To address this, we add uniform nois to each energy sample: \( \tilde{y}_i = y_i + \epsilon_i \), where \( \epsilon_i \sim \mathcal{U}(-\delta/2, \delta/2) \) and \( \delta \) is the smallest nonzero difference between sorted unique energy values~\cite{coles2001introduction, haan2006extreme, faranda2014extreme, Faranda_2013}. This smoothing step allows EVT to more accurately model the distribution of extreme values.

Given a reference energy value \( y_{\text{ideal}} \) obtained from classical Simulated Annealing (SA), we estimate how many quantum executions (each with \( s \) shots) are required to achieve, with confidence level \( \alpha \), at least one result no worse than \( y_{\text{ideal}} \). Letting \( p = \mathbb{P}(y \leq y_{\text{ideal}}) \) denote the probability that a single quantum execution yields a satisfactory result (as determined via the fitted GEV distribution), the required number of executions \( n_{\mathrm{EVT}} \) satisfies:
\[
1 - (1 - p)^{n_{\mathrm{EVT}}} \geq \alpha \quad \Rightarrow \quad n_{\mathrm{EVT}} \geq \frac{\log(1 - \alpha)}{\log(1 - p)}.
\]
The total number of required quantum shots is then simply \( n_{\mathrm{EVT}} \times s \). $s$ denotes the number of shots executed to compute a single extreme point $y_i$.

\section{EXPERIMENT}

\subsection{Experiment Setup}
\subsubsection{Quantum Devices and Algorithm Configuration}

Experiments are conducted on two gate-based quantum devices via AWS Braket: \emph{Ankaa-3} (superconducting) and \emph{Forte Enterprise 1} (ion-trap), and a simulator that is also offered by AWS Braket to be consistent with these platforms
\footnote{\href{https://aws.amazon.com/braket/quantum-computers/}{AWS Quantum Device}}. The simulator performs ideal (noise-free) quantum circuit emulation using the native gate sets of each corresponding quantum device. We solve QUBO instances using the Quantum Approximate Optimization Algorithm (QAOA) with a circuit depth of \(p=3\), optimized using COBYLA~\cite{powell1994direct}. All QAOA parameters are first tuned on a simulator, from which the best-performing set is transferred to real quantum devices. In real device, all experiments use 2000 shots. Additional platform-specific implementation details are available in the Braket SDK documentation
\footnote{\href{https://amazon-braket-sdk-python.readthedocs.io/en/stable/_apidoc/modules.html}{Braket SDK Docs}}.

\subsubsection{Datasets and Base model}

We use two standard click-through-rate (CTR) prediction datasets: Avazu and Criteo~\cite{avazu2015ctr,criteo2014}. To match quantum hardware limitations, we construct feature subsets of sizes \emph{10, 13, 15,} and \emph{18}. Feature selection is guided by XGBoost importance scores: the top 3 and bottom 2 features are retained, and the remaining are sampled from importance-ranked groups to ensure diversity. QUBO matrices \(Q\) are derived from a base recommendation  mode (DeepFM~\cite{guo2017deepfm}) following the PDQUBO framework. We also impose cardinality constraints on the number of selected features (as shown in Eq.~\ref{eq:QUBO_penalty}), set to \emph{8, 11, 12,} and \emph{14} for each subset size, respectively. Our code is available at \url{https://github.com/jiayangniu/Recsys25_QEVT}

\begin{table}[htp]
\centering
\caption{Estimated executions \( n_{\mathrm{EVT}} \) needed for quantum simulators and Quantum Processing Unit (QPU) to match or surpass classical SA at 95\% and 90\% confidence, across various feature selection scales}
\label{tab:table1}
\begin{tblr}{
  rowsep = 0pt,
  colspec = {Q[65]Q[62]Q[160]Q[160]Q[160]Q[160]},
  rows = {font=\small},
  width = \linewidth,
  cells = {c},
  cell{1}{1} = {c=2}{0.128\linewidth},
  cell{1}{3} = {c=2}{0.324\linewidth},
  cell{1}{5} = {c=2}{0.324\linewidth},
  cell{2}{1} = {c=2}{0.128\linewidth},
  cell{3}{3} = {c=4}{0.642\linewidth,c},
  cell{4}{1} = {r=3}{},
  cell{7}{1} = {r=3}{},
  cell{10}{1} = {r=3}{},
  cell{13}{1} = {c=2}{0.128\linewidth},
  vline{1,3,5,7} = {-}{},
  vline{1,2,5,7} = {13}{},
  hline{1-2,4,13,17} = {-}{},
  hline{3} = {3-6}{},
}
Dataset       &       & Avazu             &                 & Criteo            &                 \\
simulator     &       & Ankaa-3           & Forte           & Ankaa-3           & Forte  \\
Scale         & shots & 95\% /90\% >= SA    &                 &                   &                \\
10            & 500   & 7/6               & 13/10           & 6/5               & 10/8            \\
              & 1000  & 5/4               & 9/7             & 4/3               & 6/5             \\
              & 2000  & 4/3               & 5/4             & 5/4               & 4/3             \\ \hline
13            & 500   & 23/16             & 24/16           & 15/12             & 24/19           \\
              & 1000  & 13/11             & 16/13           & 7/6               & 13/11           \\
              & 2000  & 8/6               & 9/7             & 5/4               & 7/6             \\ \hline
15            & 500   & 62/47             & 103/80          & 62/48             & 70/54           \\
              & 1000  & 30/25             & 71/54           & 34/26             & 36/28           \\
              & 2000  & 19/15             & 46/35           & 22/17             & 31/24           \\
QPU           &       & Ankaa-3           & Forte           & Ankaa-3           & Forte           \\
10            & 500   & 16/13             & 25/19           & 23/17             & 31/24           \\
13            & 500   & 27/21             & 68/52           & 51/39             & 97/75           \\
15            & 500   & 76/58             & 111/87          & 114/88            & 209/161         
\end{tblr}
\end{table}

\begin{figure*}[htb]
  \centering
  \begin{subfigure}[t]{0.245\textwidth}
    \centering
    \includegraphics[width=\linewidth]{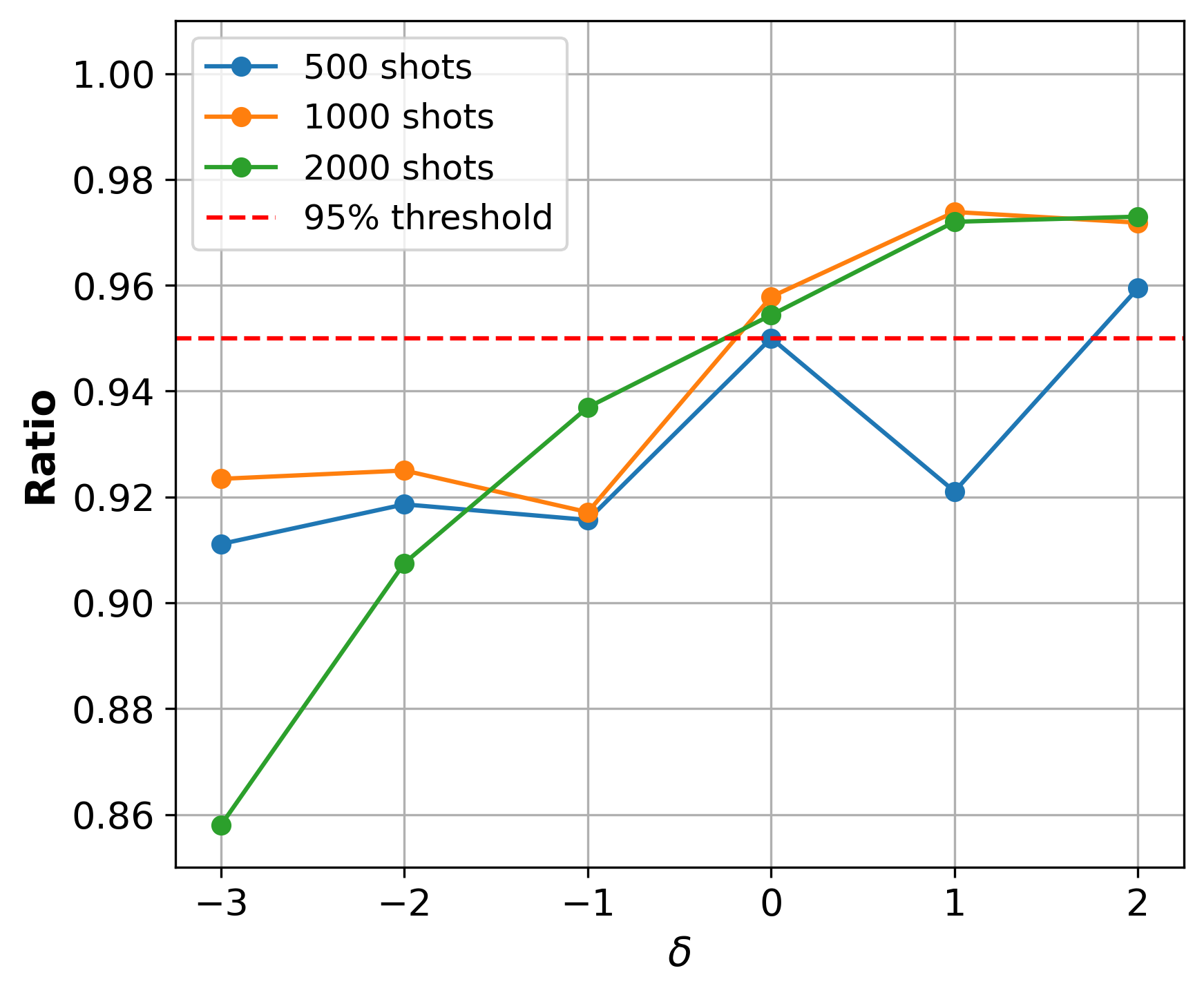}
    \caption{Avazu (13 features)}
    \label{fig:Avazu_13}
  \end{subfigure}\hfill
  \begin{subfigure}[t]{0.245\textwidth}
    \centering
    \includegraphics[width=\linewidth]{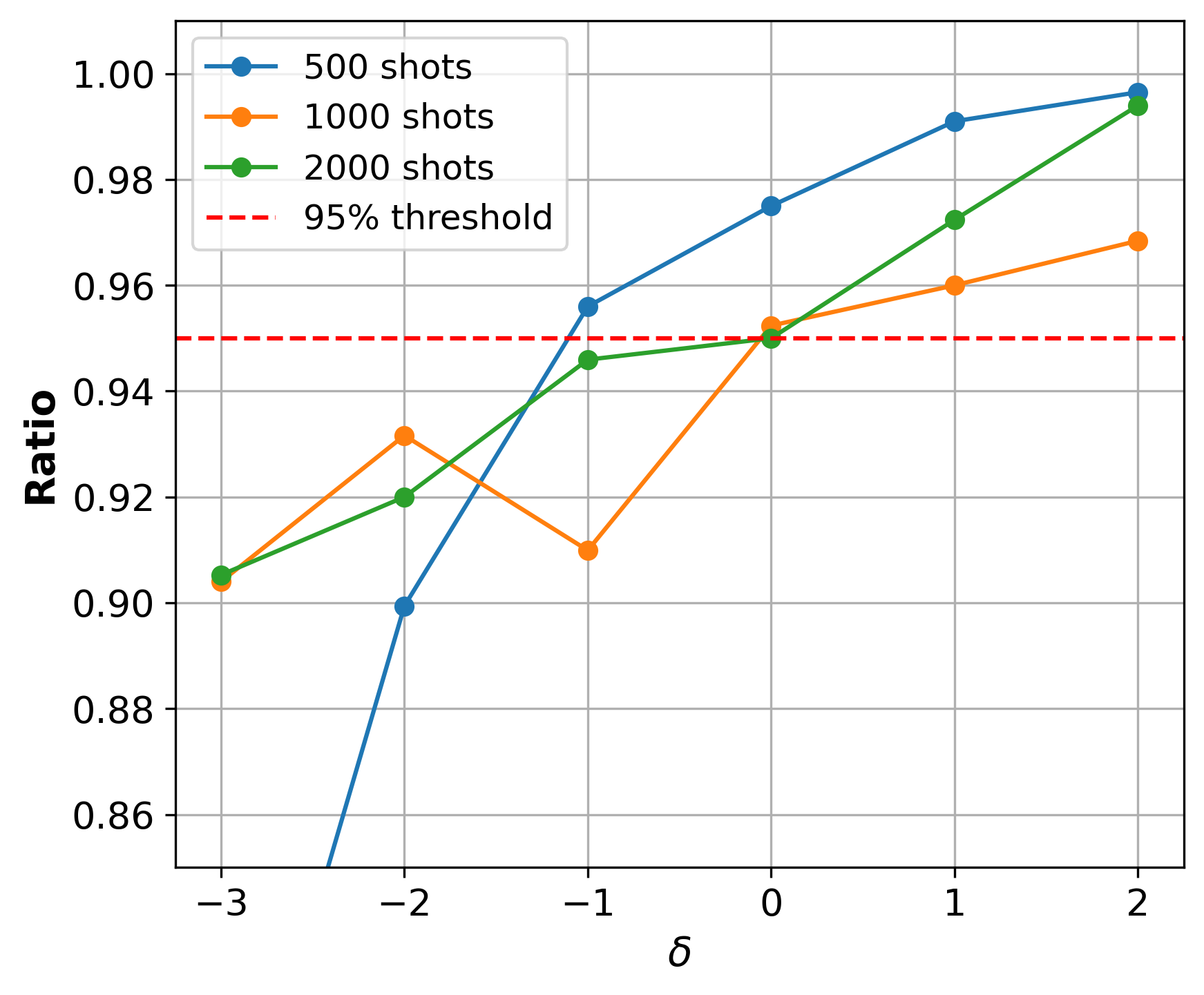}
    \caption{Criteo (13 features)}
    \label{fig:Criteo_13}
  \end{subfigure}\hfill
  \begin{subfigure}[t]{0.49\textwidth}
    \centering
    \includegraphics[width=\linewidth]{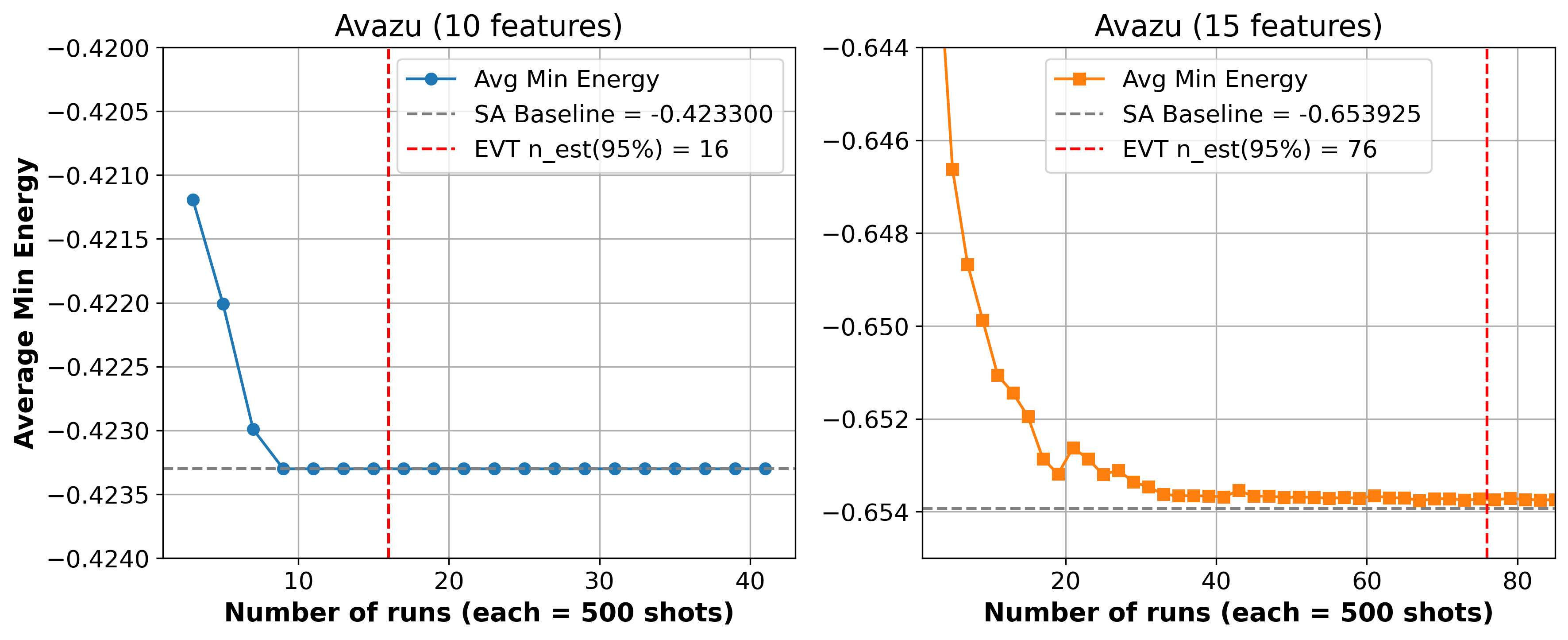}
    \caption{Average min.\ energy vs.\ runs (500 shots each) for scales 10 \& 15}
    \label{fig:avazu_runs_comparison}
  \end{subfigure}
  \caption{%
  (a) and (b) Evaluation on simulators, and (c) Evaluation and trend on physical quantum processor Ankaa-3. Note the red dashed line denotes the 95\% confidence target.}
  \label{fig:all_in_one_row}
\end{figure*}

\subsection{Experiment Results}
% \subsubsection{Preliminary Experiment:}
% The preliminary experiment compares the solving capabilities of the QAOA algorithm executed on two quantum devices against the Simulated Annealing (SA) algorithm on a classical computer.
% After optimizing the QAOA parameters, we run each quantum simulator and real quantum device with 1000 shots, and record the lowest energy observed among the 1000 shots, along with the corresponding feature subset that achieves this minimum energy. The results are presented in Table~\ref{tab:table1}.

\subsubsection{EVT-Based Shot Estimation}

We conduct experiments on the Avazu and Criteo datasets using feature subset sizes of \([10, 13, 15]\). We compare against classical simulated annealing (SA) since prior QUBO studies have commonly used SA as a performance baseline for quantum optimization~\cite{niu2024performance,ferrari2022towards}. For each configuration, we run quantum simulations under shot settings of \([500, 1000, 2000]\) to estimate the number of runs required to reach classical SA performance using EVT. On real quantum hardware (Ankaa-3 and Forte), we execute only the 500-shot setting for selected scales. The required number of extreme samples to fit the GEV distribution is estimated using the procedure described in Section~\ref{sec:eres}, which suggests a threshold of 70–80 samples for stable estimation. To ensure robustness, we conservatively collect 200 extreme samples per configuration. Each sample corresponds to the minimum energy obtained from a single quantum run. These values are then fitted to a GEV model, which is used to estimate the total number of quantum executions needed to exceed SA performance with 95\% and 90\% confidence levels.

Table~\ref{tab:table1} presents the estimated number of runs (blocks of shots) required for quantum simulators and real quantum devices to match or exceed classical simulated annealing (SA) performance at 95\% and 90\% confidence levels across different feature selection scales. A clear trend emerges: as the feature selection scale increases, more runs are required. For instance, on the Avazu dataset using the Ankaa-3 simulator with 500 shots per run, the required number of runs under 95\% confidence grows from 7 at scale 10 to 62 at scale 15. Similarly, the Forte simulator requires 13 and 103 runs, respectively, over the same scales. Additionally, increasing the number of shots per run reduces the required number of runs for a fixed scale. For example, at scale 13 on Avazu, the Ankaa-3 simulator requires 23, 13, and 8 runs when using 500, 1000, and 2000 shots, respectively. Since each configuration was independently sampled, this monotonic reduction supports the consistency and reliability of EVT-based estimation. Finally, real quantum devices require more runs than simulators, likely due to noise. On Criteo at scale 15 with 500 shots, the Ankaa-3 simulator requires 62 runs for 95\% confidence, while the physical Ankaa-3 device requires 114 runs. This gap reflects real hardware imperfections and reinforces the need for robust estimation in practical quantum optimization.

\subsubsection{Evaluation on Simulator}

To test whether our EVT model accurately predicts the number of quantum executions required to match or exceed classical SA performance, we conducted $1{,}000$ independent QAOA runs on the Avazu and Criteo datasets at a fixed feature scale of 13. For each dataset, we used three per‐run shot budgets $s\in\{500,1000,2000\}$ and measured, for each total‐shot count $(n_{\mathrm{EVT}}+\delta)\times s$, the fraction of runs whose minimal energy fell at or below the SA baseline, which is defined as our empirical “ratio”. Then, we plotted these ratios against $\delta$ from the corresponding Q-EVT run count $n_{\mathrm{EVT}}$ in Figures~\ref{fig:Avazu_13} and~\ref{fig:Criteo_13}.
Specifically, an accurate prediction implies that the curve crosses the horizontal dashed line precisely at $\delta=0$. 
Indeed, for both the 500-shot and 2000-shot settings, the empirical curves intersect at $\delta\approx0$, confirming that our EVT estimates pinpoint the required number of repetitions with high precision. While the 1000-shot configuration on Criteo exhibits slightly more scatter, its crossing remains very close to $\delta=0$, demonstrating that the method remains robust even under realistic device noise and sampling variability.

\subsubsection{Evaluation on Physical Quantum Processors}

We evaluate this by demonstrating the trend of optimisation results while increasing the number of runs on a real quantum processor.  
Using the Ankaa-3 superconducting QPU, we solve the Avazu feature‐selection problem at two fixed scales (13 and 15 features). For each scale, we sweep the per-run shot budget \(s\) from 500 to 20000 in steps of 1000. At each \(s\), we perform 20 independent QAOA executions, record the minimum energy observed in each execution, and plot the average of these 20 minima.  
Figure~\ref{fig:avazu_runs_comparison} shows that, for both scales, the average best‐observed energy steadily approaches the classical SA baseline (dashed line) as \(s\) increases. In particular, small shot budgets (below 5000) yield noticeably higher energies, while beyond \(\sim10\,000\) shots the averaged minimum energy nearly converges to the SA reference. These results demonstrate that, even on noisy gate-based quantum hardware, simply allocating more shots per run can substantially improve the probability of finding high-quality solutions.

\subsubsection{Limitations and Breakdown Scenarios}

While our EVT-based framework provides accurate shot estimates across a variety of settings, we observe two scenarios where the approach may begin to break down. First, when the quantum algorithm consistently yields identical optimal outputs within a small number of shots, the lack of output variability limits the statistical basis required for reliable EVT modeling. Second, when quantum outputs remain significantly suboptimal even after many executions, the estimated shot count may become unstable or overestimated due to poor alignment with the classical solution landscape.

\section{Conclusion}
Our results demonstrate that Extreme Value Theory (EVT) provides an effective framework for estimating the number of quantum executions (runs) required to match or exceed classical simulated annealing (SA) performance. 
Across both quantum simulators and real gate-based devices, the predictions of Q-EVT align with empirical outcomes at various feature selection scales. 
This allows for robust, confidence-based shot estimation without requiring exhaustive empirical tuning. 

\begin{acks}
This research is supported by the ARC Discovery Project (DP2101007- 43) and RACE (RMIT Advanced Cloud Ecosystem). We would like to express our sincere gratitude to RACE for providing access to quantum computing hardware and computational infrastructure, which was essential for conducting our experiments and advancing this research.
\end{acks}

\bibliographystyle{ACM-Reference-Format}
\bibliography{ref}
\end{document}